\documentclass[preprint,showkeys,showpacs,preprintnumbers,amsmath,amssymb]{revtex4}

\usepackage{graphicx}  
\usepackage{dcolumn}   
\usepackage{bm}        

\newcommand{\ve}[1]{{\bf\underline #1}}
\newcommand{\D}{\partial}
\newcommand{\vve}[1]{{\bf\underline{\underline #1}}}

\newcommand{\Dt}[1]{\frac {\D #1} {\D t}}

\newcommand{\dt}[1]{\frac {d #1} {d t}}
\newcommand{\delt}{{\Delta t}}
\newcommand{\hot}{O(\delt^2)}

\newcommand{\DI}[1]{\frac {\D #1} {\D x_1}}	
\newcommand{\DII}[1]{\frac {\D #1} {\D x_2}}

\newcommand{\Dij}[1]{\frac {\D^2 #1} {\D x_i \D x_j}}
\newcommand{\DIDI}[1]{\frac {\D^2 #1} {\D x_1^2}}
\newcommand{\DIIDII}[1]{\frac {\D^2 #1} {\D x_2^2}}

\newcommand{\excl}[1]{{\backslash \hspace{-0.3em} #1}}

\newcommand{\bracket}[1]{\left[#1\right]}

\newcommand{\parenth}[1]{\left(#1\right)}

\DeclareSymbolFont{AMSb}{U}{msb}{m}{n}
\DeclareMathSymbol{\R}{\mathbin}{AMSb}{"52}

\begin{document}
\preprint{Preprint submitted to {\it Phys. Rev. Lett.}}

        \title{Information flow within stochastic dynamical systems}
        \author{X. San Liang}
        \email{sanliang@cims.nyu.edu}
        \affiliation{Courant Institute of Mathematical Sciences \\
                        New York, New York 10012}

        \date{\today}

\begin{abstract}

Information flow or information transfer is an important concept in dynamical 
systems which has applications in a wide variety of scientific disciplines.
In this study, we show that a rigorous formalism can be established 
in the context of a generic stochastic dynamical system.
The resulting measure of of information transfer possesses a property of
transfer asymmetry and, when the stochastic perturbation to the receiving
component does not rely on the giving component, has a form same 
as that for the corresponding deterministic system. 
An application with a two-dimensional system is presented, 
and the resulting transfers are just as expected. A remarkable
observation is that, for two highly correlated time series, 
there could be no information transfer from one certain series, 
say $x_2$, to the other ($x_1$). That is to say, the evolution of $x_1$
may have nothing to do with $x_2$, even though $x_1$ and $x_2$ 
are highly correlated.
Information transfer analysis thus extends the traditional notion
of correlation analysis by providing a quantitative measure of 
causality between time series.

\end{abstract}




 \maketitle

Information transfer, or information flow as it is called,
is an important concept in dynamical systems and general physics
which has been of interest since decades 
ago\cite{history}-\cite{Majda}.
Practical applications have been reported in fields like
neuroscience\cite{neuro} and atmosphere-ocean science\cite{weather_infoflow},
and are envisioned in the diverse disciplines such as turbulence research,
material science, nanotechnology, to name a few, where 
ensemble forecasts 
are involved and predictability 
becomes an issue.
Recently, Liang and Kleeman\cite{LK05}\cite{LK07} put
this important concept on a rigorous footing 
in the context of deterministic dynamical systems.
In this study, we will show that a rigorous formulation can also 
be obtained when the dynamical system is stochastic.
We consider only two-dimensional (2D) systems; systems of
higher dimensionality will be reported elsewhere.\cite{LK08}

We start with a brief review of the work in \cite{LK05} 
to educe the strategy for the building of our formalism 
for stochastic systems. Consider a 2D system
	\begin{eqnarray}	
	\dt {\ve x} = \ve F (\ve x,t), \label{eq:govd}
	\end{eqnarray}
where $\ve F=(F_1, F_2)$, and the state variables
$\ve x = (x_1, x_2) \in \R^2$. The randomness is limited within the 
initial condition. For notational simplicity, we do not distinguish random 
variables and deterministic variables, which should be clear in the
context. (In probability theory, they are usually distinguished with
lower and upper cases.) Let $\rho$ be the joint probability density
of $x_1$ and $x_2$, and suppose that it and its derivatives
have compact support.
Without loss of generality, consider the information
transfer from $x_2$ to $x_1$. We need the marginal density of $x_1$,
$\rho_1(t; x_1) = \int_\R \rho dx_2$, and the marginal (Shannon) entropy,
$H_1=-\int_\R \rho_1 \log\rho_1\ dx_1$.
$H_1$ varies as the system moves forward. 
Its variation is due to two different mechanisms, one due to $x_1$ itself, 
written as $\dt {H_1^*}$, another due to the transfer from $x_2$.
The latter is the very information transfer, which we will write as
$T_{2\to1}$ hereafter.
The rate of information transfer from $x_2$ to $x_1$
is therefore the difference between $\dt {H_1}$ and $\dt {H_1^*}$,
$T_{2\to1} = \dt {H_1} - \dt {H_1^*}$.
Among the terms on the right hand side, $\dt {H_1}$ can be derived
from the Liouville equation\cite{Lasota} corresponding to 
 (1);
the key is the derivation of $\dt {H_1^*}$, 
the entropy change as $x_1$ evolves
on its own. In \cite{LK05}, this is achieved with the aid of 
a theorem established therein: 
The joint entropy of $(x_1,x_2)$, $H = -\iint_{\R^2}\rho\log\rho\ d\ve x$, 
evolves as
        \begin{center}
        \framebox[0.3\textwidth]{
        \begin{minipage}[c]{1\textwidth}
	\begin{eqnarray}	\label{eq:dH}
	\dt H = E(\nabla\cdot \ve F).
	\end{eqnarray}
        \end{minipage}
        }
        \end{center}
Here the operator $E$ is the mathematical expectation with respect to 
	$\rho$.
Liang and Kleeman then intuitively argued that
	\begin{eqnarray}	\label{eq:dH1star}
	\dt {H_1^*} = E\parenth{\DI {F_1}},
	\end{eqnarray} 
a result later on they rigorously proved,\cite{LK07}
and hence obtained the transfer $T_{2\to1}$.

The above formalism has been generalized to the information transfer 
within a deterministic system of arbitrary dimensionality\cite{LK07};
the key equation (\ref{eq:dH1star}) has also been used
to form the transfer between two subspaces\cite{Majda}.
The generalization, however, encounters difficulty
when stochasticity is involved.
Consider a system
	\begin{eqnarray} 	\label{eq:gov}
	d {\ve x} = \ve F(\ve x, t) dt + \vve B(\ve x, t) d{\ve w},
	\end{eqnarray}
where $\ve w = (w_1,w_2)$ is a standard 2D Wiener process 
($\dt{{\ve w}}$ ``white noise''), $\vve B=\parenth{b_{ij}}$ 
the perturbation amplitude.
There is no such elegant form as (\ref{eq:dH})
for the evolution of $H$. One thus cannot obtain 
$\dt{H_1^*}$ intuitively as (\ref{eq:dH1star}) is obtained.

But on the other hand, $\dt {H_1^*}$ may be equally understood as the 
rate of change of the marginal entropy of $x_1$ 
with the effect from $x_2$ excluded. This alternative interpretation,
as we used in \cite{LK07}, sheds light on the above problem.
To reflect this interpretation, we will denote the term as 
$\dt {H_{1\excl2}}$ henceforth, the subscript $\excl2$ signifying
``$x_2$ excluded''. The rate of information transfer 
from $x_2$ to $x_1$ is thence
	\begin{eqnarray}	\label{eq:T21}
	T_{2\to1} = \dt {H_1} - \dt {H_{1\excl2}}.
	\end{eqnarray}
Here the key issue is how to find $\dt {H_{1\excl2}}$,
which we will show shortly after the evaluation of $\dt {H_1}$.

Entropy evolution is related to density evolution.
Corresponding to (\ref{eq:gov})
there is a Fokker-Planck equation:\cite{Lasota}
	\begin{eqnarray}	\label{eq:fokker}
	\Dt\rho + \DI{(F_1\rho)} + \DII{(F_2\rho)}
	= \frac 1 2 \sum_{i,j=1}^2 \Dij{(g_{ij}\rho)},
	\end{eqnarray}
where 
	$g_{ij} = g_{ji} = \sum_{k=1}^2 b_{ik} b_{jk},$
$i,j=1,2$. This integrated over $\R$ with respect to $x_2$ gives
the evolution of $\rho_1$:
	\begin{eqnarray}	\label{eq:fokker1}
	\Dt{\rho_1} + \int_\R \DI {(F_1\rho)} dx_2
		= \frac 1 2 \int_\R \DIDI {(g_{11} \rho)} dx_2.
	\end{eqnarray}
Note in the derivation we have used the fact that $\rho$ and its
derivatives vanish at the boundaries as they are compactly supported.
For notational succinctness, we will henceforth suppress
the integral domain $\R$, unless otherwise noted.
Multiplying (\ref{eq:fokker1}) by $-(1+\log\rho_1)$ followed by an integration
with respect to $x_1$ over $\R$, one obtains
	\begin{eqnarray*}
	\dt {H_1} - \iint \log\rho_1 \DI {(F_1\rho)}\ dx_1 dx_2
	= -\frac 1 2 \iint \log\rho_1 \DIDI {(g_{11}\rho)}\ dx_1dx_2.
	\end{eqnarray*}
Integrating by parts, this is reduced to
	\begin{eqnarray}	\label{eq:dH1}
	\dt {H_1} = - E\parenth{F_1 \DI {\log\rho_1}}
		    - \frac 1 2 E\parenth{g_{11} \DIDI {\log\rho_1}},
	\end{eqnarray}
where $E$ stands for expectation with respect to $\rho$.

The key part of this study is the evaluation of $H_{1\excl2}$.
Examine a small time interval $[t, t+\delt]$. $H_{1\excl2}$ is 
the time rate of change of the marginal entropy of $x_1$ as $x_2$ 
frozen as a parameter instantaneously at $t$. 
So one needs to consider a system on $[t,t+\delt]$ suddenly modified at 
time $t$ from that prior to $t$.
Clearly,
$H_{1\excl2}$ cannot be derived from the Fokker-Planck equation 
(\ref{eq:fokker1}), where the dynamics is consistent through time.
One has to go back to the definition of derivative to achieve the goal.
Let the marginal entropy evolved from $t$ to $t+\delt$ with $x_2$
frozen at $t$ be $H_{1\excl2}(t+\delt)$. We then have
	$$\dt {H_{1\excl2}} = 
	  \lim_{\delt\to0} \frac {H_{1\excl2}(t+\delt) - H_1(t)} \delt,$$
and the whole problem now boils down to the derivation of 
$H_{1\excl2}(t+\delt)$. 
In \cite{LK07}, we discretize the deterministic equation 
	(1)
and evaluate the Frobenius-Perron operator for the discretized system
to compute the modified marginal entropy. 
For the stochastic system (\ref{eq:gov}), however, 
there is no such a simple operator.
We need a different approach for the problem.

Denote by $x_{1\excl2}$ the first component after $x_2$ is fixed as a parameter.
The stochastic system (\ref{eq:gov}) is changed to
	\begin{eqnarray}
	&&d {x_{1\excl2}} = F_1(x_{1\excl2}, x_2, t) dt + \sum_k b_{1k} d{w_k},
				\  {\rm on}\ [t, t+\delt],  
				\label{eq:gov_no2}		\\
	&&x_{1\excl2} = x_1	\qquad {\rm at\ time}\ t.
	\end{eqnarray}
Correspondingly the density $\rho_{1\excl2}$ evolves following
the following Fokker-Planck equation 
	\begin{eqnarray}
	&& \Dt {\rho_{1\excl2}} + \DI {(F_1\rho_{1\excl2})}
		= \frac 1 2 \DIDI {(g_{11}\rho_{1\excl2})},
			\qquad t\in[t, t+\delt]
			\label{eq:fokker1_no2} \\
	&& \rho_{1\excl2} = \rho_1\qquad {\rm at}\ t,
	\end{eqnarray}
where $g_{11} = \sum_k b_{1k}^2$.
Recall by definition, the Shannon entropy may be understood as 
the expectation of a function of the state variable formed by 
minus logarithm composite with its density. 
This motivates one to introduce a function 
of $x_1$, $f_t(x_1) = \log\rho_{1\excl2}(t, x_1)$,
whose evolution is obtained by dividing (\ref{eq:fokker1_no2}) 
by $\rho_{1\excl2}$:
	\begin{eqnarray*}
	\Dt {f_t} + \frac 1 {\rho_{1\excl2}} \DI {F_1\rho_{1\excl2}}
	= \frac 1 {\rho_{1\excl2}} \DIDI {g_{11}\rho_{1\excl2}}.
	\end{eqnarray*}
In a discretized version, this is
	\begin{eqnarray*}
	f_{t+\delt}(x_1) = f_t(x_1) - \frac \delt{\rho_1} \DI {(F_1\rho_1)}
		    + \frac \delt {2\rho_1} \DIDI {(g_{11}\rho_1)} + \hot,
	\end{eqnarray*}
where the fact
$\rho_{1\excl2} = \rho_1$ at time $t$ has been used. 
(Functions without arguments explicitly written out are supposed to 
be evaluated at $x_1(t)$.)
So
	\begin{eqnarray*}
	f_{t+\delt} (x_{1\excl2}(t+\delt))
	=  f_t(x_{1\excl2}(t+\delt)) 
	   - \frac\delt{\rho_1} \DI{(F_1\rho_1)}
	   + \frac \delt {2\rho_1} \DIDI {(g_{11}\rho_1)} + \hot.
	\end{eqnarray*}
The $x_{1\excl2}(t+\delt)$ in the argument can be expanded by the 
Euler-Bernstein approximation\cite{Lasota} of (\ref{eq:gov_no2}):
	\begin{eqnarray*}
	x_{1\excl2}(t+\delt) = x_1(t) + F_1 \delt + \sum_k b_{1k} 
			\Delta w_k + h.o.t.
	\end{eqnarray*}
Substituting back and performing Taylor series expansion,
we get
	\begin{eqnarray}		\label{eq:expect}
	f_{t+\delt} (x_{1\excl2}(t+\delt))
	&=& 
	  f_t\parenth{x_1 + F_1\delt + \sum_k b_{1k} \Delta w_k}
	  - \frac\delt{\rho_1} \DI{(F_1\rho_1)}
	  + \frac \delt {2\rho_1} \DIDI {(g_{11}\rho_1)} + \hot		\cr
	&=&
	  f_t(x_1) 
	     + \DI {f_t} \parenth{F_1\delt + \sum_k b_{1k} \Delta w_k}
	     + \frac 1 2 \DIDI {f_t} \parenth{F_1\delt 
	     + \sum_k b_{1k} \Delta w_k}^2 				\cr
	& &\qquad
	  - \frac\delt{\rho_1} \DI{(F_1\rho_1)}
	  + \frac \delt {2\rho_1} \DIDI {(g_{11}\rho_1)} + \hot.
	\end{eqnarray}
Take expectation on both sides, the left hand side is $-H_{1\excl2}(t+\delt)$,
and the first term on the right hand side is $-H_1(t)$. Note that
$\Delta w_k \sim N(0, \delt)$ for a Wiener process $w_k$. So
	\begin{eqnarray*}
	E \Delta w_k = 0, \qquad E(\Delta w_k)^2 = \delt.
	\end{eqnarray*}
The second term on the r.h.s. is 
	\begin{eqnarray*}
	\delt \cdot E\parenth{F_1 \DI{f_t}} + 
	E \parenth{\DI{f_t} \sum_k b_{1k} \Delta w_k}
	= \delt\cdot E\parenth{F_1 \DI{f_t}},
	\end{eqnarray*}
where we have used the fact that $\Delta w_k$ is independent of $(x_1, x_2)$, 
and hence expectation can be taken inside directly with $\Delta w_k$, which
eliminates
	$E \parenth{\DI{f_t} \sum_k b_{1k} \Delta w_k}$.
For the same reason, the third term after expansion leaves only one sub-term of
order $\delt$, namely,
	\begin{eqnarray*}
	&& \frac 1 2 E\bracket{\DIDI{f_t} 
		\sum_k b_{1k}\Delta w_k \sum_j b_{1j}\Delta w_j} \cr
	&& = \frac 1 2 E\bracket{\DIDI{f_t}\parenth{
		\sum_k b_{1k}^2 (\Delta w_k)^2 +
		\sum_{k\ne j} b_{1k} b_{1j} \Delta w_k \Delta w_j}}.
	\end{eqnarray*}
Recall that the perturbations are independent.
The summation over $k$$\ne$$j$ inside the parentheses
thus vanishes after expectation is performed.
The first summation is equal to $g_{11} \delt$,
by the definition of $g_{ij}$ and the fact $E(\Delta w_k)^2 = \delt$.
So the whole term is $\frac \delt 2 E\bracket{g_{11} \DIDI{f_t}}$.
With all these put together, expectation of (\ref{eq:expect}) gives
(note $f_t = \log\rho_{1\excl2}(t;x_1) = \log\rho_1$)
	\begin{eqnarray*}
	&& H_{1\excl2}(t+\delt) 
	=  H_1(t)
	- \delt\cdot E\parenth{F_1 \DI {\log\rho_1}}
	- \frac \delt 2 E\parenth{g_{11} \DIDI {\log\rho_1}} \cr
	&& 
	+ \delt\cdot E\parenth{\frac 1 {\rho_1} \DI {(F_1\rho_1)} }
	- \frac \delt 2 E\parenth{\frac 1 {\rho_1} \DIDI{(g_{11}\rho_1)}}
	+ \hot.
	\end{eqnarray*}
The second and fourth terms on the right hand side can be combined to give
	$ \delt\cdot E\parenth{\DI{F_1}}.$
So	
	\begin{eqnarray}
        \dt {H_{1\excl2}}	\label{eq:dH1_no2}
	&=& \lim_{\delt\to0} \frac {H_{1\excl2}(t+\delt) - H_1(t)} \delt \cr
	&=& E\parenth{\DI {F_1} }
	 - \frac 1 2 E\parenth{g_{11}\DIDI{\log\rho_1}}
	 - \frac 1 2 E\parenth{\frac 1 {\rho_1} \DIDI{(g_{11}\rho_1)}}.
	\end{eqnarray}
In the equation, the second and the third terms on the right hand side
are from the stochastic perturbation. The first term is precisely 
(\ref{eq:dH1star}), the key result obtained in \cite{LK05} through 
intuitive argument based on the theorem (\ref{eq:dH}).
The above derivation supplies a proof of this argument.

The information transfer from $x_2$ to $x_1$ is obtained
by subtracting (\ref{eq:dH1_no2}) from (\ref{eq:dH1}):
	\begin{eqnarray}	\label{eq:transfer}
	T_{2\to1} 
		  &=& - E\parenth{F_1 \DI {\log\rho_1}} 
		      - E\parenth{\DI{F_1}}
	      + \frac 1 2 E\parenth{\frac 1 {\rho_1} \DIDI {(g_{11}\rho_1)}} \cr
	 &=& - E\parenth{\frac 1 {\rho_1} \DI{(F_1\rho_1)}}
	     + \frac 1 2 E\parenth{\frac 1 {\rho_1} \DIDI{(g_{11}\rho_1)}},
	\end{eqnarray}
where $E$ is the expectation with respect to $\rho(x_1,x_2)$.
Notice that the conditional density of $x_2$ on $x_1$, $\rho_{2|1}$,
is $\rho / \rho_1$. If we write the expectation with
respect to $\rho_{2|1}$ as $E_{{2|1}}$, 
the above formula may be further simplified:
	\begin{center}
	\framebox[0.6\textwidth]{
	\begin{minipage}[c]{1\textwidth}
	\begin{eqnarray}	\label{eq:transfer1}
	\ \ \ T_{2\to1} 
	 = - E_{2|1} \parenth{\DI{(F_1\rho_1)}}
	     + \frac 1 2 E_{2|1} \parenth{\DIDI{(g_{11}\rho_1)}}.
	\end{eqnarray}
	\end{minipage}
	}
	\end{center}
This is the transfer from $x_2$ to $x_1$. Likewise, the transfer from $x_1$ to
$x_2$ can be obtained:
	\begin{eqnarray}	\label{eq:transfer2}
	T_{1\to2} 
	 = - E_{1|2} \parenth{\DII{(F_2\rho_2)}}
	     + \frac 1 2 E_{1|2} \parenth{\DIIDII{(g_{22}\rho_2)}},
	\end{eqnarray}
where $\rho_2 = \int\rho\ dx_1$ is the marginal density of $x_2$.

Among the two terms of (\ref{eq:transfer1}) the first is the same in form 
as the information transfer obtained in \cite{LK05} for the corresponding
deterministic system.
The contribution from the stochasticity that modifies the formula 
is in the second term.  
An interesting observation is that, if $g_{11} = \sum_k b_{1k}^2$ 
is independent of $x_2$, this term vanishes.
To see this, notice that $\int \rho_{2|1} dx_2 = 1$, which results in
	\begin{eqnarray*}
	E_{2|1} \parenth{\DIDI {(g_{11}\rho_1)}}
	= \int \DIDI {(g_{11}\rho_1)}\ dx_1 = 0.
	\end{eqnarray*}
Wee thus have the following property:
	\begin{itemize}
	\item[\ ]
	{\it
	Given a stochastic system component,  
	if the stochastic perturbation is independent of another 
	component, then the information transfer from the latter is 
	the same in form as that for the corresponding deterministic system.
	}
	\end{itemize}
This property is interesting since a large proportion of noise appearing 
in real problems are additive, that is to say, $b_{ij}$, 
and hence $g_{ij}$, are often constant.
This theorem shows that, in terms of information transfer, 
these stochastic systems function like deterministic.
But, of course, the similarity is just in form; 
they are different in value. 
The first part on the right hand side of (\ref{eq:transfer1}) 
actually has stochasticity embedded in the marginal density.

Another property is the concretization of the requirement of transfer
asymmetry emphasized by Schreiber:\cite{Schreiber}
	\begin{itemize}
	\item[\ ]
	{\it If the evolution of $x_1$ is independent 
	of $x_2$, then $T_{2\to1}$ is zero.
	}
	\end{itemize}
In fact, if neither $F_1$ and $g_{11}$ have dependency on $x_2$, 
the integrals in (\ref{eq:fokker1}) can be evaluated and the whole equation 
becomes a Fokker-Planck equation for $\rho_1$. In this case, 
$x_1$ behaves like an independent variable. 
So by intuition, there should be no information flowing from $x_2$.
This is indeed true by formula (\ref{eq:transfer1}).
If $F_1 = F_1(x_1)$, integration can be made for $\rho_{2|1}$ 
with respect to $x_2$ inside the double integral, giving a zero $T_{2\to1}$.

The formulas (\ref{eq:transfer1}) and (\ref{eq:transfer2}) are expected to be
applicable in a wide variety of fields. To demonstrate an application, 
consider a 2D linear system:		
	\begin{eqnarray}	\label{eq:langevin}
	\dt {\ve x} = \vve A\ \ve x dt + \vve B d{\ve w}, 
	\end{eqnarray}
where $\vve A = (a_{ij})$ and $\vve B = (b_{ij})$ are constant matrices.
Further suppose that $\ve x$ has an initial Gaussian distribution;
it is then Gaussian all the time\cite{Gardiner}, with
a mean $\ve\mu = (\mu_1, \mu_2)^T$ and a covariance matrix $\vve C = (c_{ij})$ 
evolving as
	\begin{subequations}	\label{eq:mean}
	\begin{eqnarray}
	 d{\ve\mu} / dt &=& \vve A\ \ve\mu,	\\
	d {\vve C} / dt &=& \vve A\ \vve C + \vve C\ \vve A^T + \vve B\ \vve B^T.
	\end{eqnarray}
	\end{subequations}
The solution of these equations determines the density 
	$$\rho(\ve x) = \frac 1 { 2\pi |\vve C|^{1/2}}
	e^{-\frac 1 2 {(\ve x - \ve \mu)}^T {\vve C}^{-1} (\ve x - \ve\mu)}$$
which after substituted into (\ref{eq:transfer1}) and (\ref{eq:transfer2}) 
gives the transfers between $x_1$ and $x_2$.

\begin{figure} [h]
   \begin{center}
   \includegraphics[angle=0,width=1.\textwidth]
    {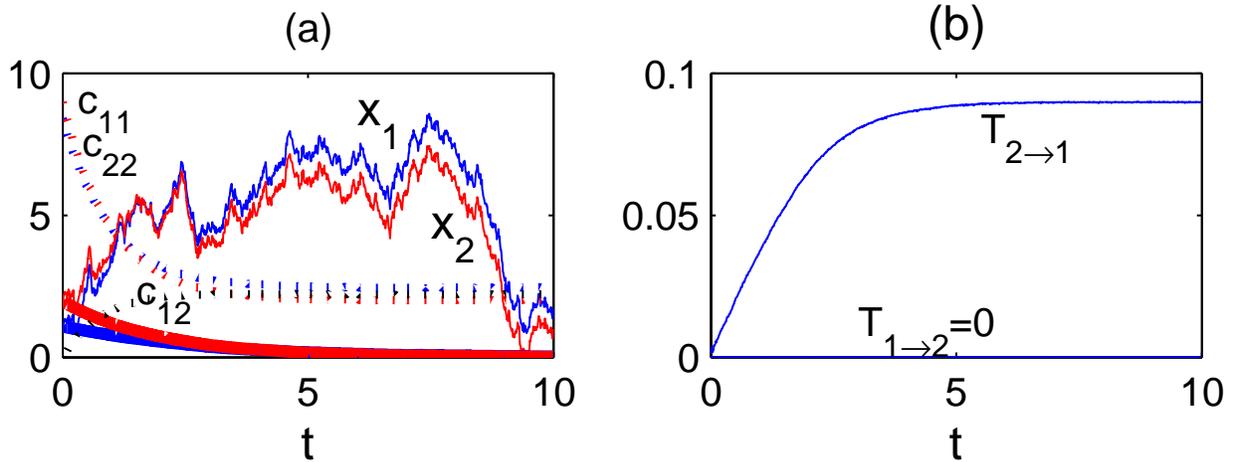}
   \caption
        { 
	(a) A solution of (\ref{eq:mean}) with $a_{21}=0$:
	$\ve\mu$ (thick solid), $\vve C$ (dotted).
	Also shown is a sample path (solid) starting from $\ve\mu(0)$.
	(b) The computed information transfers $T_{2\to1}$ (upper) 
	and $T_{1\to2}=0$.}
        \protect{\label{fig:sys1}}
   \end{center}
   \end{figure}

For an example,
let all the entries of $\vve B$ be 1, and 
$a_{11}=a_{22}=-0.5$, $a_{12}=0.1$,
leaving $a_{21}$ open for experiment.
First consider $a_{21}=0$. It is easy to show that this system 
has an equilibrium solution:
	$\ve\mu = (0,0)$, 
	$c_{11} = 2.44$, $c_{12}=c_{21}=2.2$, 
	$c_{22} = 2$,
whatever the initial conditions are.
Fig.~\ref{fig:sys1}a shows the time evolutions of $\ve\mu$ and $\vve C$
initialized with $\ve\mu(0) = (1,2)$  and 
$c_{11}(0)=c_{22}(0)=9$, $c_{12}(0)=c_{21}(0)=0$;
also shown is a sample path of $\ve x$ starting from $\ve\mu(0)$.
In this system, 
$F_2 = -0.5 x_2$ has no dependence on $x_1$, 
and $g_{ij} = \sum_k b_{ik} b_{jk}$ are all constants,
so $T_{1\to2} = 0$ by the property established above.
The computed result confirms this inference. 
In Fig.~\ref{fig:sys1}b, $T_{1\to2}$ is zero through time.
The other transfer, $T_{2\to1}$, increases monotically and eventually
approaches a constant. 

An interesting observation about the typical sample path 
in Fig.~\ref{fig:sys1}a is the high correlation between 
$x_1$ and $x_2$, in contrast to the zero information transfer
$T_{1\to2}$. That is to say, even though $x_1(t)$ and $x_2(t)$ are highly
correlated, the evolution of $x_2$ has nothing to do with $x_1$.
Through this simple example one sees how 
information transfer extends the traditional notion of 
correlation analysis and/or mutual information analysis
by including causality\cite{Granger}.

In the second experiment, we let $a_{21}=0.1=a_{12}$, resulting in a system
symmetric between $x_1$ and $x_2$. One thus naturally expects two transfers
equal in value. The computed results show that this is indeed so.
The transfer $T_{2\to1}$ is equal to $T_{1\to2}$ (not shown).
(If $\mu_1 \ne \mu_2$, initially they may be different, but merge together
soon after the transient period.) 
In the third experiment, $a_{21} = 0.2 > a_{12}$;
the influence of $x_1$ on $x_2$ is larger than
that of $x_2$ on $x_1$, so one expects a larger $T_{1\to2}$ than
$T_{2\to1}$. Again, the computed result agrees with the inference 
(not shown). The formulas (\ref{eq:transfer1}) and (\ref{eq:transfer2})
are verified with this example.

We have rigorously established a formalism of information transfer
within 2D stochastic dynamical systems, which is measured by the
rate of entropy transferred from one component to another.
The measure possesses a property of transfer asymmetry and, 
when the stochastic perturbation to the receiving component does 
not rely on the giving component, has a form same
as that for the corresponding deterministic system. An application
with a linear system has been presented, from which
one sees that correlation does not necessarily mean causality; 
for two highly correlated time series, the one-way information 
transfer could be zero.
Information transfer provides a quantitative way of establishing 
the causal relation between dynamical events. 
This quantification of causality is expected to have important 
applications in a wide variety of scientific disciplines.

The author has benefited from several important scientific discussions 
with Richard Kleeman on this subject. He also read through an early version 
of the manuscript, and his comments are greatly appreciated.



\end{document}